\documentclass[a4paper,11pt]{article}
\pdfoutput=1 

\usepackage{jheppub}
\usepackage{subfig}
\usepackage{graphicx}
\usepackage{epsfig}
\usepackage{amsmath, amsthm, amssymb}
\usepackage{amsfonts}
\usepackage{color}
\usepackage{xspace}
\usepackage[dvipsnames]{xcolor}
\usepackage{paralist}
\usepackage{multirow}

\preprint{MS-TP-20-41, KA-TP-22-2020, P3H-20-081, ZU-TH-58/20}
\title{Electroweak \boldmath$t \bar t$ hadroproduction in the presence of heavy  \boldmath$Z'$ and \boldmath$W'$ bosons at NLO QCD in POWHEG}

\author[a,c]{Mohammad Mahdi Altakach,}
\author[b]{Tom\'{a}\v{s} Je\v{z}o,}
\author[c]{Michael Klasen,}
\author[d]{Jean-Nicolas Lang,}
\author[a]{Ingo Schienbein}

\affiliation[a]{Laboratoire de Physique Subatomique et de Cosmologie,
 Universit\'e Grenoble-Alpes, CNRS/IN2P3,
 53 Avenue des Martyrs, 38026 Grenoble, France}

\affiliation[b]{Institut f\"ur Theoretische Physik, Karlsruhe Institut f\"ur Technologie, 76128 Karlsruhe, Germany}

 \affiliation[c]{Institut f\"ur Theoretische Physik, Westf\"alische
 Wilhelms-Universit\"at M\"unster, Wilhelm-Klemm-Stra\ss{}e 9, 48149
 M\"unster, Germany}

\affiliation[d]{Physik-Institut, Universit\"at Z\"urich, 8057 Z\"urich, Switzerland}

\emailAdd{altakach@lpsc.in2p3.fr}
\emailAdd{tomas.jezo@kit.edu}
\emailAdd{michael.klasen@uni-muenster.de}
\emailAdd{jlang@physik.uzh.ch}
\emailAdd{ingo.schienbein@lpsc.in2p3.fr}

\abstract{
We extend and improve upon our previous calculation 
of electroweak top-quark pair hadroproduction in extensions
of the Standard Model with extra heavy neutral and charged spin-1 
resonances.
In particular, we allow for flavour-non-diagonal $Z'$ couplings and take into account non-resonant production 
in the SM and beyond including the contributions with $t$-channel $W$ and $W'$ bosons. All amplitudes are generated using the Recola2 package.
As in our previous work, we include NLO QCD corrections and consistently match to parton showers with the 
POWHEG method fully taking into account the interference effects between SM and new physics amplitudes. 
We consider the Sequential Standard Model, the Topcolour model, as well as the Third Family Hypercharge Model featuring 
non-flavour-diagonal $Z'$ couplings which has been proposed recently to explain the anomalies in $B$ decays.
We present numerical results for $t \bar t$ cross sections at hadron colliders with a centre-of-mass energy up to 100 TeV.
}


\begin{document} 
\maketitle
\flushbottom

\clearpage
\section{Introduction}
\label{sec:intro}

The Standard Model (SM) of particle physics, based on an SU(3)$_C \times$SU(2)$_L
\times$ U(1)$_Y$ gauge symmetry, is an extremely successful theory that accounts
for a wide range of experimental measurements at both the intensity and energy
frontiers. Nevertheless, it is widely believed to be incomplete for different
reasons. On the observational side, the SM does not include gravity, it does not
provide a candidate for a cold dark matter particle, the CP violating phase of
the SM CKM matrix is not sufficient to explain the matter--anti-matter asymmetry
observed in the Universe, and massive neutrinos are, a priori, not accounted for
in the SM. In addition, the SM and in particular its scalar sector
suffer from a variety of naturalness problems raising the questions
why CP-violation in the strong interaction is absent or at least strongly suppressed
and why the Higgs boson mass is stable under quantum corrections.
On the aesthetical side, it is not clear why
the SM gauge group is a direct product of three independent symmetry factors, 
why there are three generations of fermions, why their masses span several orders of magnitude, 
and what gives rise to the complicated pattern in the CKM and PMNS mixing matrices.
The general expectation has therefore been for a long time that close to the
electroweak scale new physics beyond the SM should be present, that allows one to
at least partially understand the structure of the SM and avoid fine-tuning its
26 free parameters.

Despite the fact that no signals of new physics have been found in the first two
runs of the Large Hadron Collider (LHC) at CERN, there are still high hopes that
new particles will show up in the future high-luminosity runs of the LHC. It is
also clear that such signals will likely appear as small deviations from the SM
predictions, which makes precision calculations of both the SM background and the
new physics signals increasingly important.
The focus of this paper is on models with new heavy, electrically charged or
neutral spin-one resonances, usually denoted by $W'$ and $Z'$, respectively. Such
resonances are predicted by several well-motivated extensions of the SM, e.g.\
Grand Unified Theories, theories with a new strong interaction at the TeV
scale, or models with large extra dimensions, and are extensively sought after by
three of the experimental collaborations (ATLAS, CMS and LHCb) at the LHC. In this
respect it is noteworthy that $Z'$ models with a non-universal flavour structure
\cite{Allanach:2018lvl,Allanach:2019iiy}, where the $Z'$ couples differently to
the fermions of the three SM families, are viable candidates to explain the
current $B$-flavour anomalies \cite{Aaij:2017vbb,Aaij:2015oid,Aaij:2013qta,%
Aaij:2017vad,Aaboud:2018mst,CMS:2014xfa,Chatrchyan:2013bka,Bobeth:2017vxj,%
Khachatryan:2015isa}.

In many cases, the strongest constraints on the parameter space of models with
$Z'$ and $W'$ resonances come from searches with dilepton final states
\cite{Accomando:2013sfa,Accomando:2019ahs}. In this case, precise predictions
at next-to-leading order (NLO) of QCD including the resummation of soft gluon
terms at next-to-leading logarithmic accuracy can be obtained with the
{\tt Resummino} code \cite{Fuks:2007gk,Fuks:2013vua}, which has been used
previously to derive limits on $Z'$ and $W'$ masses using data for dilepton
final states \cite{Jezo:2014wra,Klasen:2016qux} and to provide predictions for
the high-energy/high-luminosity options of the LHC \cite{CidVidal:2018eel}.
However, top-quark observables are also very interesting  since the third
generation plays a prominent role in the SM due to the large Yukawa coupling of
the top quark. Therefore, it is quite conceivable that new gauge bosons,
similarly to the Higgs boson, couple predominantly to the top quark
\cite{Basso:2012sz}.  In 2015, some of us performed a calculation of
next-to-leading order QCD corrections to the electroweak (EW) $t\bar{t}$
production in the presence of a $Z'$ resonance \cite{Bonciani:2015hgv}. The
calculation properly accounted for the interference between SM and new physics
amplitudes in a semi-automated fashion. It was implemented in the {\tt
POWHEG\,BOX} framework \cite{Nason:2004rx,Frixione:2007vw,Alioli:2010xd}, that
allows for a consistent matching of the fixed NLO calculation with Parton
Shower (PS) Monte Carlo generators. The NLO+PS results obtained with this tool,
dubbed {\tt PBZp}, are useful since they bridge the gap between
first-principles higher-order calculations and the complex detector signatures
and data of the experimental community. The {\tt PBZp} code is therefore
regularly used in $Z'$ searches by the ATLAS and CMS collaborations
\cite{Aaboud:2018mjh,Aaboud:2019roo,%
Aad:2020kop,CMS:2016zte,CMS:2016ehh,Sirunyan:2017uhk,CMS:2017mhy}.

In this article, we perform a complete re-calculation of the processes
implemented in the code {\tt PBZp} including a number of improvements:
\begin{inparaenum}[(i)]
\item The amplitudes have been calculated using the {\tt Recola2} package
  \cite{Denner:2017wsf}. This package has been designed to automate the
  calculations of amplitudes in theories beyond the SM including QCD and
  electroweak corrections at NLO. Our calculation is one of the first
  to use and validate this tool for a BSM calculation. Implementing the
  amplitudes obtained with {\tt Recola2} into Monte Carlo event generators, here
  within the {\tt POWHEG\,BOX} framework is an important aspect since it makes
  the tool more useful for the LHC experiments.
\item The new code can now deal with flavour-non-diagonal and generation
  non-universal couplings such that the $Z'$-models mentioned above explaining
  the $B$-flavour anomalies can be implemented.
\item Another new feature is that the calculation now includes $t$-channel $W$
  and $W'$ contributions. We study their importance numerically as a function of
  both the new gauge boson mass and the collider energy and propose useful
  kinematic cuts to disentangle the different contributions.
\item As before, all interference terms are fully taken into account, and the
  photon induced channels for the SM are included, properly matched within
  POWHEG. As discussed in Ref.\ \cite{Bonciani:2015hgv} the latter can give a
  sizable contribution to the cross section.
\end{inparaenum}
This work paves the way for a similar calculation of the NLO QCD corrections to
the process $p p \to W/W'/Z' \to t\bar{b}$ using {\tt Recola2} for the generation of
the amplitudes and including a proper implementation within the {\tt POWHEG\,BOX}
framework.

The remainder of this paper is organised as follows: In Sec.\ \ref{sec:hadroproduction}
we define the production of top-quark pairs at hadron colliders including new electroweak
gauge bosons, focusing on the perturbative organisation of the cross section and its
contributions at leading and next-to-leading order in the strong and electroweak coupling
constants. In Sec.\ \ref{sec:calculation} we describe our calculation with a focus on
the aspects that differ from our previous calculation in Ref.\ \cite{Bonciani:2015hgv}. 
This discussion should be useful for other {\tt Recola2}-based calculations
in the future. Next, in Sec.\
\ref{sec:models} we summarise the models for which we present numerical results in
Sec.\ \ref{sec:numerics}. In addition to a study of the effect of cuts we will show
results for fiducial cross sections for
a range of heavy resonance masses and different centre-of-mass energies. Here, the
purpose is to present the new features of the {\tt PBZp} code instead of aiming at
exhaustive phenomenological studies of these models which we leave for future work.
Finally, in Sec.\ \ref{sec:conclusions} we present a summary and our conclusions.

\section{Hadroproduction of top-quark pairs}
\label{sec:hadroproduction}

The cross section for the hadroproduction of a $t \bar t$ pair, $AB \to t \bar t X$, is given by a convolution of the
parton distribution functions (PDFs) inside the two incoming hadrons [$f_{a/A}(x_a,\mu_F)$, $f_{b/B}(x_b,\mu_F)$] 
with the short distance cross sections [$d\hat \sigma_{ab}$]:
\begin{eqnarray}
d\sigma & = & \sum_{ab}\int f_{a/A}(x_a,\mu_F)f_{b/B}(x_b,\mu_F)\,d\hat \sigma_{ab}(\mu_R,\mu_F) dx_a dx_b\quad \, .
\label{eq:hadro}
\end{eqnarray}
Here, $\mu_F$ and $\mu_R$ are the factorisation and renormalisation scales, respectively, and a sum over all relevant
partonic channels, $ab \to t \bar t X$ is performed. This sum depends on various details like the heavy flavour scheme,
the perturbative order, and the model. 
Here we work in a 5-Flavour Number Scheme (5-FNS) including all relevant contributions with
$u$, $d$, $s$, $c$, $b$ (anti-)quarks, gluons and unless explicitly stated also photons in the initial state.

Up to next-to-leading order the hard scattering cross sections have the following perturbative expansion in the  strong ($\alpha_s$)  
and electroweak ($\alpha$) coupling constants:
\begin{eqnarray}
\hat \sigma &=& 
\hat \sigma_{2;0}(\alpha_{S}^2)
 + \hat \sigma_{3;0}(\alpha_S^3)
 + \hat \sigma_{2;1}(\alpha_S^2\alpha)
 + \boldsymbol{\hat \sigma_{0;2}(\alpha^2)}
+ \boldsymbol{\hat \sigma_{1;2}(\alpha_{S}\alpha^2)}
\nonumber
\\
&\phantom{=}&
 + \boldsymbol{ \hat \sigma_{1;1}(\alpha_{S}\alpha)}
 + \hat \sigma_{0;3}(\alpha^3)\,,
 \label{eq:1.2}
\end{eqnarray}
where the numerical indices $(i;j)$ in $\hat \sigma_{i;j}$ represent the powers in $\alpha_s$ and $\alpha$, respectively,
and the dependence on the renormalisation and factorisation scales and the parton flavour indices "ab" have been suppressed.
We now briefly describe the different contributions where the terms highlighted in bold have been included in our calculation:
\begin{itemize}
\item $\boldsymbol{ \hat \sigma_{0;2}, \hat \sigma_{1;2}}$:
In this paper, we focus on the tree-level electroweak top-quark pair production, $\hat \sigma_{0;2}$,
and the NLO QCD corrections to it, $\hat \sigma_{1;2}$. 
$\hat \sigma_{0;2}$ receives contributions from the $s$-channel amplitudes $q \bar{q} \to (Z',Z,\gamma) \to t \bar t$
including the $Z'$ signal and its interference with the photon and SM $Z$ boson.
Due to the resonance of the $Z'$ boson, we expect these terms to be the most relevant for new physics searches.
In addition, we include new contributions from diagrams with non-resonant $t$-channel exchange of $W$, $W'$ and $Z'$ bosons that were not 
considered in Ref.\ \cite{Bonciani:2015hgv}. 
Note that out of these, the first two take into account CKM mixing and the last one is only allowed in models with flavour-non-diagonal couplings
due to the absence of a top quark PDF in a 5-FNS. 
A particular advantage of the $\hat \sigma_{0;2}$ contribution is that the calculation of higher order QCD corrections to it, $\hat \sigma_{1;2}$, can be carried 
out in a model-independent way. 
\item $\boldsymbol{\hat \sigma_{1;1}}$: 
We also consider the term $\hat \sigma_{1;1}$ which receives contributions from the photon induced subprocess $\gamma g \to t \bar t$  
and the previously not considered interference of the $s$-channel QCD and the $t$-channel electroweak top-pair production (see below).
Note that the photon induced subprocess is needed for a consistent treatment of the mass singularities in the 
process $gq \to t\bar t q$ when the $t$-channel photon is collinear to the quark $q$. It turns out that this contribution is numerically important.
However, we neglect photon-initiated contributions to $\hat \sigma_{0;2}$ and $\hat \sigma_{1;2}$.

\item $\hat \sigma_{2;0},\hat \sigma_{3;0}$:
These terms are the contributions from the
SM QCD ``background'' processes $q\bar{q},gg\to t\bar{t}[g]$,
$q g \to t \bar{t} q$, and $\bar q g \to t \bar t \bar q$ which have been computed in the late 1980s \cite{Nason:1987xz,Nason:1989zy,Beenakker:1988bq,Beenakker:1990maa}.
Furthermore, NLO calculations for heavy quark correlations \cite{Mangano:1991jk}
and $t \bar t$ spin correlations \cite{Bernreuther:2001rq,Bernreuther:2004jv} are available too.
The terms $\hat \sigma_{2;0}$ and $\hat \sigma_{3;0}$ are not affected by the presence of $Z'$ or $W'$ bosons and 
are readily available in many NLO+PS event generators \cite{Frixione:2007nw,Campbell:2014kua,Jezo:2016ujg,Gleisberg:2008ta,Alwall:2014hca,Bellm:2015jjp}.
\item $\hat \sigma_{2;1}$: 
This term represents the electroweak corrections to the QCD backgrounds. 
Within the SM, a gauge-invariant subset was first investigated neglecting the interferences between QCD and electroweak interactions 
arising from box-diagram topologies and pure photonic contributions \cite{Beenakker:1993yr} and later also including
additional Higgs boson contributions arising in Two-Higgs-Doublet Models \cite{Kao:1999kj}.
The rest of the electroweak corrections was subsequently calculated in a series of papers and
included also $Z$-gluon interference effects and QED corrections with real and virtual photons
\cite{Kuhn:2005it,Moretti:2006nf,Bernreuther:2005is,Bernreuther:2006vg,Hollik:2007sw}. 
In principle, $\hat \sigma_{2;1}$ would also receive contributions from new $Z'$ and $W'$ resonances.
However, these contributions are expected to give a small correction to $\hat \sigma_{2;1}$
and since they are highly model-dependent due to the rich structure of the 
scalar sector in many models we don't include them in our calculation.
\item $\hat \sigma_{0;3}$:
Finally, the purely electroweak term in Eq.\ (\ref{eq:1.2}) would also receive contributions from new $Z'$ and $W'$ resonances.
It is of order ${\cal O}(\alpha^3)$ and parametrically suppressed compared to the other terms.
We therefore do not include it in our calculation.\footnote{Note, however, that despite
the parametric suppression they have been shown to be important in the SM in a region with large top transverse momentum \cite{Pagani:2016caq,Gutschow:2018tuk}.}
\end{itemize}

\section{NLO QCD corrections to electroweak top-pair production}
\label{sec:calculation}

In this section, we present our new calculation of the NLO QCD corrections to electroweak top quark-pair production
in the presence of heavy $Z'$ and $W'$ spin-one resonances where we put particular emphasis on the changes
with respect to our previous calculation in Ref.\ \cite{Bonciani:2015hgv}.

\subsection{Analysis chain}
\label{sec:chain}
In \cite{Bonciani:2015hgv}, the Feynman diagrams were generated using {\tt QGRAF} \cite{Nogueira:1991ex} and then translated into amplitudes 
using {\tt DIANA} \cite{Tentyukov:1999is}.
The calculation was carried out in the Feynman gauge in $D=4 - 2 \epsilon$ dimensions in order to regularise the
ultraviolet (UV) and infrared divergences using {\tt FORM} \cite{Vermaseren:2000nd}.
The loop integrals were then reduced to a basis of three master integrals having well-known solutions \cite{Hooft:1978xw}\footnote{These integrals are 
the massive tadpole, the equal-masses two-point function, and the massless two-point function.}
using integration-by-parts identities 
\cite{Tkachov:1981wb,Chetyrkin:1981qh} in the form of the Laporta algorithm \cite{Laporta:2001dd}
as implemented in the public tool {\tt REDUZE} \cite{2010CoPhC.181.1293S,vonManteuffel:2012np}.
Traces involving the Dirac matrix $\gamma_5$ were treated in the Larin prescription \cite{Larin:1993tq}
by replacing $\gamma_\mu \gamma_5 = i \frac{1}{3!} \epsilon_{\mu\nu\rho\sigma} \gamma^\nu \gamma^\rho \gamma^\sigma$.
We used the on-shell scheme to subtract the UV divergences.
Furthermore, in order to restore the Ward identities and thus preserve gauge invariance at one loop, an additional finite renormalisation for vertices involving $\gamma_5$
was performed. Finally, we verified analytically that the remaining soft and soft-collinear divergences cancel in the sum of the real and virtual contributions, as a consequence of the
Kinoshita-Lee-Nauenberg (KLN) theorem, explicitly using the Catani-Seymour subtraction~\cite{Catani:2002hc}\footnote{Note, however, that
the actual treatment of the soft and collinear divergences in {\tt PBZp} was (and still is) done numerically within POWHEG using the 
Frixione-Kunszt-Signer (FKS) subtraction \cite{Frixione:1995ms}.}.
 
We now use the public library {\tt Recola2} to generate the amplitudes for the models in Sec.\ \ref{sec:models}.
{\tt Recola2} is an extension of {\tt Recola} \cite{Actis:2016mpe} for the computation of tree and one-loop amplitudes in the Standard Model
and beyond. In {\tt Recola}, one-loop amplitudes are decomposed in terms of tensor coefficients and
tensor integrals, the latter being model independent and evaluated with the help
of the {\tt Collier} tensor integral library \cite{Denner:2016kdg}.
Thus the model-dependent part only concerns the tensor coefficients and
rational parts of type $R_2$ \cite{Ossola:2008xq} that are being constructed with the help of a
{\tt Recola2} model file in a recursive and numerical way. 
The model file used in our study has been generated using the
toolchain {\tt FeynRules} \cite{Alloul:2013bka} and {\tt Rept1l}
\cite{Denner:2017vms} and is publicly available.\footnote{The model file can be found on the official website \url{https://recola.hepforge.org/} under model files.}
{\tt Recola} regularises amplitudes in dimensional regularisation with
space-time dimension ${\rm D} = 4 - 2\epsilon$, adopting by default the 
{\tt COLLIER} normalisation for 1-loop integrals \cite{Denner:2016kdg}.
More precisely, conventional dimensional regularisation is used which treats all
particles and momenta in $\rm D$ dimensions. Similarly, the Lorentz algebra is
upgraded to $\rm D$-dimensions with a special treatment of 
$\gamma_5$, known as naive dimensional regularisation (NDR) \cite{Denner:2019vbn}.
As the treatment of $\gamma_5$ is delicate in 
${\rm D} \ne 4$ dimensions, it can be formulated as a problem of determining the
correct rational term of type $R_2$ which may not have a $\gamma_5$ scheme
dependence or would otherwise prohibit defining (chiral) gauge symmetry
of the theory in a consistent way. In NDR, rational terms for amplitudes with closed fermion loops
and external vector bosons are evaluated using a reading point prescription,
giving up on the cyclicity of the trace. This procedure guarantees that no
symmetries of the theory are being broken, at least to one-loop order.
The UV renormalisation is carried out in the complete on-shell scheme for all
particles \cite{Denner:1991kt}.
Since in this work we start from electroweak production including QCD
corrections, no further renormalisation of couplings is required. 
Finally, with POWHEG we use the Binoth Les Houches Accord
conventions \cite{Alioli:2013nda} in which the UV renormalised 1-loop amplitude is given as the
following Laurent series \cite{Binoth:2010xt}\footnote{This requires calling {\tt
set\_delta\_ir\_rcl(0,$\pi^2/6$)} in {\tt Recola}.}:
\begin{align}
  \mathcal{A}_1 = C(\epsilon)
  \left(
    \frac{A_2}{\epsilon^2}
    +
    \frac{A_1}{\epsilon}
    +
    \frac{A_0}{\epsilon}
  \right), \quad
  \text{with} \quad
  C(\epsilon) = \frac{4\pi}{\Gamma(1-\epsilon)}\, .
\end{align}

One final technical point that is noteworthy concerns the treatment of Goldstone bosons.
{\tt Recola2} performs computations in Feynman gauge that requires, in general,
the inclusion of Goldstone bosons which drop out whenever a Goldstone boson is attached
to a massless quark line. However, for the processes under consideration and in
the presence of flavour mixing this is no longer the case.
For this reason we include Goldstone bosons associated to $W'$ and $Z'$ that mimic
the interaction of the SM Goldstone bosons in such a way that the amplitudes
are equivalent to a computation in unitary gauge. 

\subsection{Parton level processes}
\label{sec:processes}

In the following, we will describe the amplitudes contributing to $\hat \sigma_{1;1}$, $\hat \sigma_{0;2}$ and $\hat \sigma_{1;2}$  in more detail.
As already mentioned, we work in a 5-Flavour Number Scheme  including all relevant contributions with massless $u, d, s, c, b$ (anti-)quarks, gluons
and photons in the initial state.
Furthermore, we allow for a completely general flavour structure for the couplings of the $Z'$ and $W'$ bosons to the Standard Model
fermions.

\subsubsection{Leading-order contributions}

\begin{figure}[ht]
\centering
\subfloat{\includegraphics[width=\textwidth]{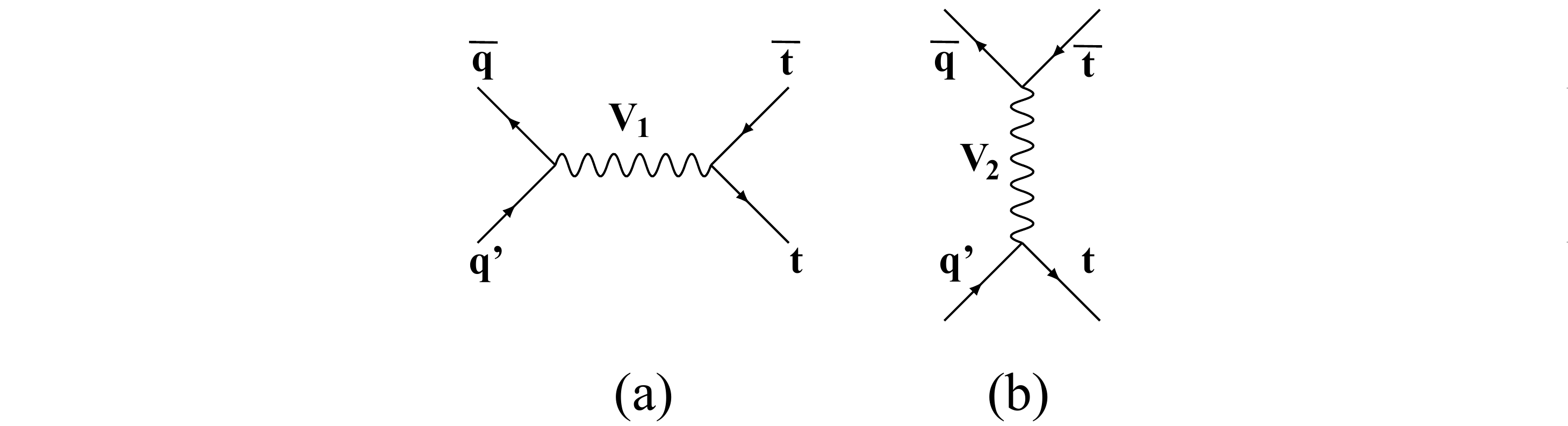}}
\caption{Born amplitudes $A_{0;1}$ contributing to the electroweak top-pair production 
cross section $\hat \sigma_{0;2}$.
a) $s$-channel contribution where $V_1 \in \{\gamma, Z, Z' \}$ and $q, q' \in \{u, d, s, c, b\}$.
In the case of $Z'$-exchange we allow for flavour-changing amplitudes with $q' \ne q$
and we don't show the corresponding amplitudes with a charge-conjugated initial state.
b) $t$-channel contributions where $V_2 \in \{W,W', Z' \}$.
All allowed amplitudes with the same initial (and final) state are added coherently.}
\label{fig:born-pbzp-diagrams}
\end{figure}

The Born amplitudes $A_{0;1}$ contributing to the electroweak top-pair production cross-section $\hat \sigma_{0;2} \sim |A_{0;1}|^2$
are shown in Fig.\ \ref{fig:born-pbzp-diagrams}.

Figure  \ref{fig:born-pbzp-diagrams}a) depicts in a compact form the contribution with an $s$-channel vector boson $V_1$ which can be
a photon, a SM $Z$-boson, or a new heavy $Z'$-boson. In the previous version of  \texttt{PBZp} we allowed
only for flavour-diagonal $Z'$-couplings with $q'=q \in \{u,d,s,c,b\}$. In this calculation we include contributions with 
flavour-changing couplings of the $Z'$ boson to the quarks in the initial state 
($q' \bar q  \in \{u \bar u,c \bar u, d \bar d, s \bar d, b \bar d, d \bar s, s \bar s, b \bar s, u \bar c, c \bar c, d \bar b, s \bar b, b \bar b \}$).
Note that the corresponding amplitudes with a charge-conjugated initial state are not shown.
In the new calculation, we now also take into account the $t$-channel  diagrams in Fig.\ \ref{fig:born-pbzp-diagrams}b)
where the exchange boson $V_2$ can be a $W$, $W'$, or $Z'$ boson. 
Needless to say that those amplitudes in Fig.\ \ref{fig:born-pbzp-diagrams}a) and b) having the same initial (and final) states
are added coherently.

As was discussed in \cite{Bonciani:2015hgv}, the diagrams in Fig.\  \ref{fig:born-pbzp-diagrams}a) have zero interference with 
the QCD amplitude $q \bar q \to g^* \to t \bar t$ since such interference terms are proportional to the vanishing colour trace $\operatorname{Tr}(T^a)$.
On the other hand, the $t$-channel diagrams in Fig.\  \ref{fig:born-pbzp-diagrams}b) with $q'=q$ do interfere with the QCD amplitude 
$q \bar q \to g^* \to t \bar t$. 
Thus they contribute to $\hat \sigma_{1;1}$ and we take them into account.

\subsubsection{One-loop virtual corrections}

\begin{figure}[ht]
\centering
\subfloat{\includegraphics[width=\textwidth]{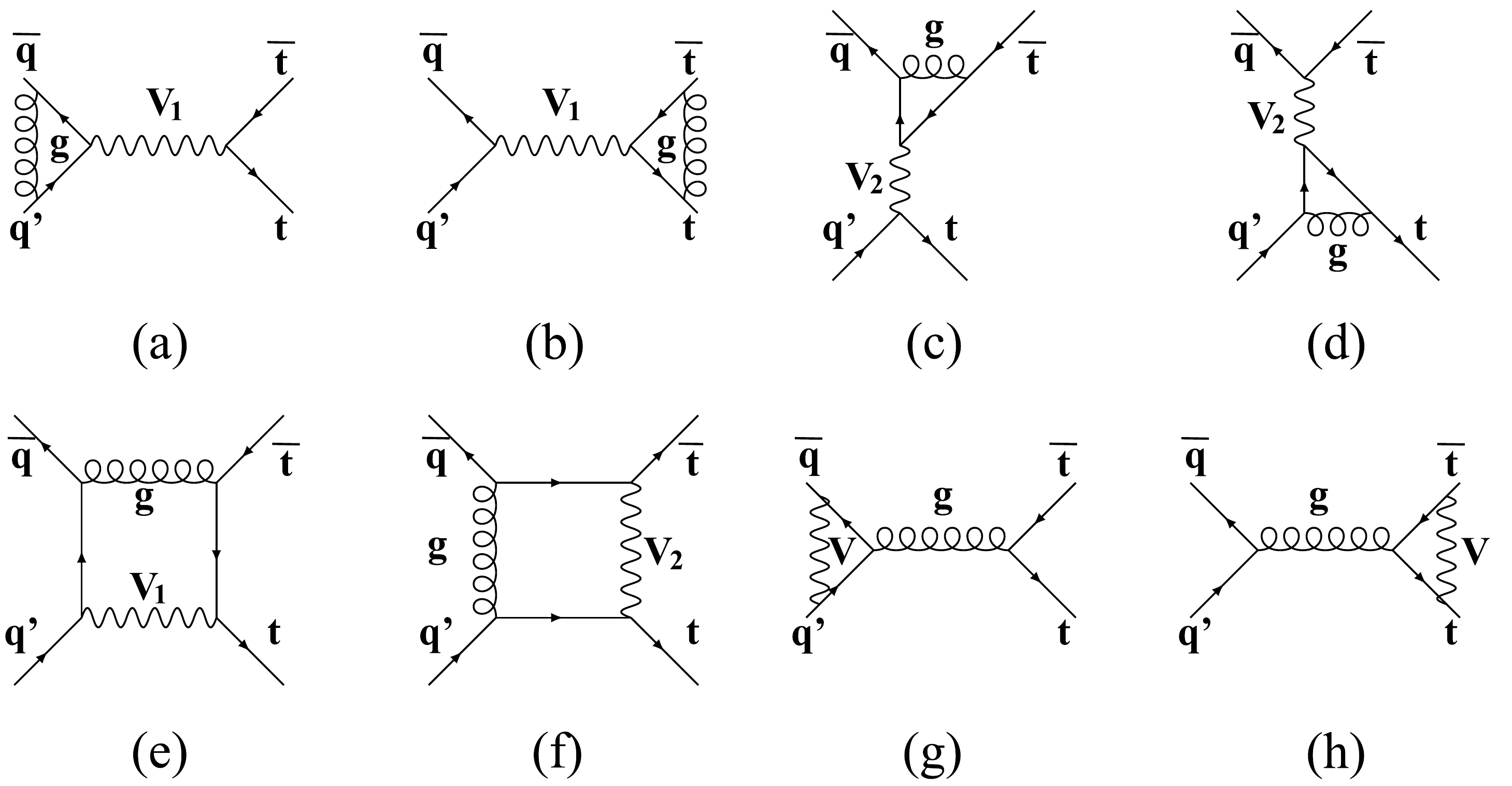}}
\caption{One-loop QCD corrections, $A_{1;1}$, to the diagrams in Fig.\ \ref{fig:born-pbzp-diagrams}.
As before, $V_1 \in \{\gamma, Z, Z' \}$, $V_2 \in \{W,W', Z' \}$, and $V=V_1 \cup V_2$ denotes
the union of $V_1$ and $V_2$.}
\label{fig:virtual-rpbzpwp-diagarams}
\end{figure}

In Fig.\ \ref{fig:virtual-rpbzpwp-diagarams} we show the one-loop QCD corrections, $A_{1;1}$, to the diagrams in Fig.\ \ref{fig:born-pbzp-diagrams}.
They contribute to the electroweak top-pair production at ${\cal O}(\alpha_s \alpha^2)$ 
due to the interference of these diagrams with the Born amplitudes
in Fig.\ \ref{fig:born-pbzp-diagrams}: $\hat \sigma_{1;2}^{\rm V} \sim 2 \Re[A_{0;1} A_{1;1}^*]$.

Note again, that there is no interference of the diagrams in Figs.\ \ref{fig:virtual-rpbzpwp-diagarams}a) and b) with the
Born level QCD amplitude $q \bar q \to g^* \to t \bar t$, whereas the diagrams in Figs.\ \ref{fig:virtual-rpbzpwp-diagarams}c) and d) 
will interfere with it contributing to $\hat \sigma_{2;1}$. As discussed in Sec.\ \ref{sec:hadroproduction}, we don't consider 
the effect of heavy new resonances on $\hat \sigma_{2;1}$ in this work.

  In our re-calculation we define the virtual corrections as the set of diagrams
  with no vector bosons inside the loops, effectively treating them as background-fields in
  {\tt Recola2} \cite{Denner:1994xt, Denner:2017vms}.  The resulting QCD loop corrections
  constitute a gauge-invariant subset which can be seen by realising that if
  either of the quark lines is replaced by an auxiliary non colour-charged
  conserved current (e.g. a lepton-lepton vector interaction), the so-defined
  virtual corrections represent not only the full QCD corrections and are thus
  gauge-independent, but, moreover, they do not depend on the specific form of
  the auxiliary conserved current. Therefore, the statement holds true for the EW
  production if one vetoes diagrams with a gluon exchange between the two
  different quark lines, i.e.\ excluding box corrections. 
  In summary, we therefore include diagrams such as
Figs.\ \ref{fig:virtual-rpbzpwp-diagarams}c) and d), but we omit
amplitudes in the second row of Fig.\ \ref{fig:virtual-rpbzpwp-diagarams}.
In principle one could compute all corrections, but the renormalisation
of the amplitudes in Figs.\  \ref{fig:virtual-rpbzpwp-diagarams}g) and h)
requires one to account for quark mixing self-energy diagrams with transitions between
different flavours due to $W$, $W'$, or $Z'$-bosons in the loop. Such quark mixing in the renormalisation is currently not
implemented in {\tt Recola2/Rept1l}.
In the limit of a diagonal CKM matrix and including only diagonal $W'$ and $Z'$
couplings, we investigated in our calculation the impact of the diagrams in 
Fig.\ \ref{fig:virtual-rpbzpwp-diagarams}e)-h) and it turns
out that in this case, their contribution is negligibly small.
Since any deviation from this “diagonal” setup is additionally suppressed by the
small off-diagonal couplings we expect the contributions from Fig.\
\ref{fig:virtual-rpbzpwp-diagarams}e)-h) to also remain negligible in this
general “non-diagonal” case.

\subsubsection{Real emission corrections}

\begin{figure}[ht]
\centering
\subfloat{\includegraphics[width=\textwidth]{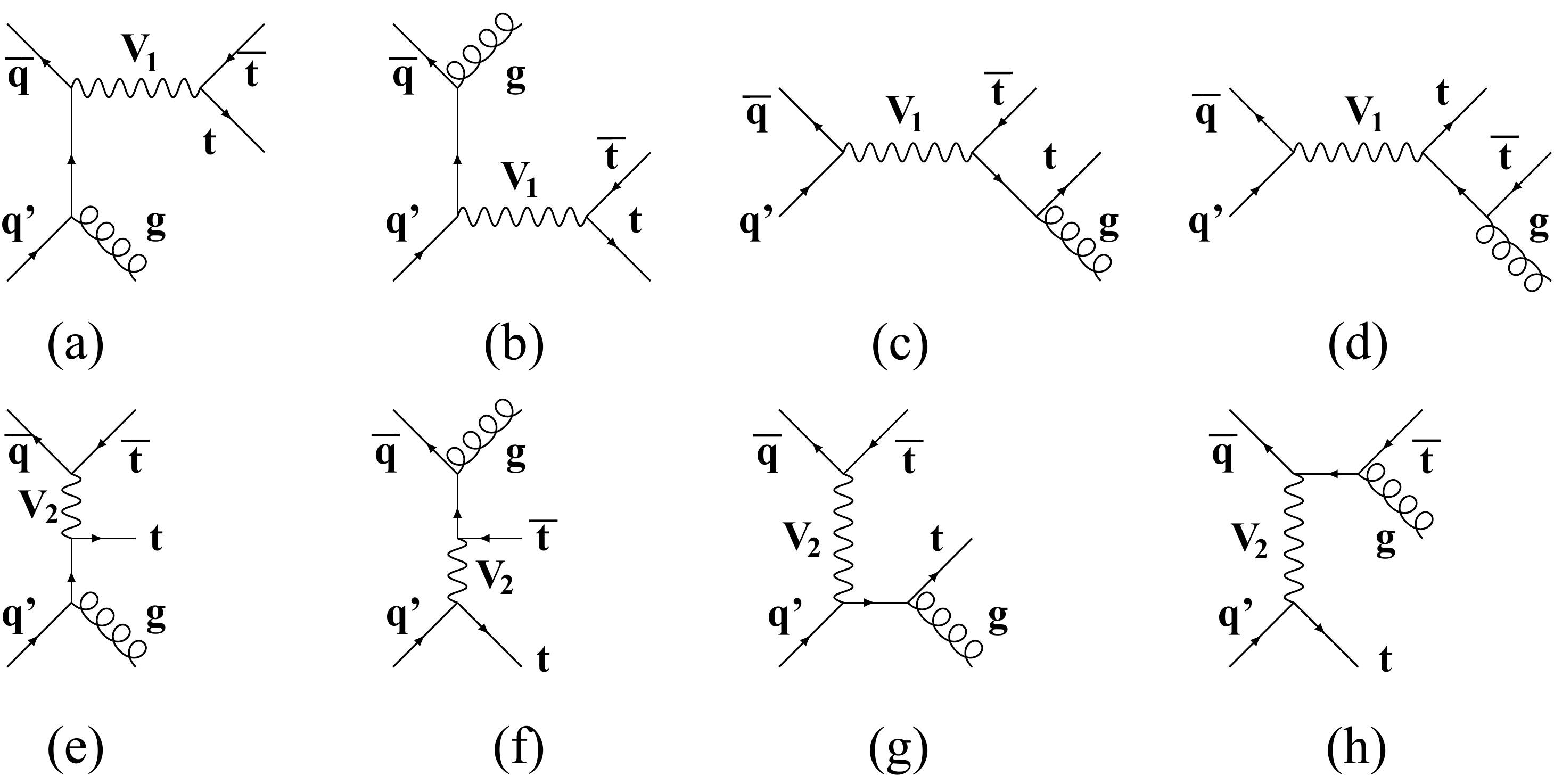}}
\caption{Diagrams contributing to  the $q' \bar q \to t \bar t g$ subprocesses at order ${\cal O}(\alpha_s \alpha^2)$.
As in Fig.\ \ref{fig:born-pbzp-diagrams}, $V_1 \in \{\gamma, Z, Z' \}$, $V_2 \in \{W,W', Z' \}$, and $q, q' \in \{u, d, s, c, b\}$.
All allowed amplitudes with the same initial (and final) state are added coherently.}
\label{fig:qq-real-pbzpwp-diagrams}
\end{figure}

\begin{figure}[ht]
\centering
\subfloat{\includegraphics[width=0.85\textwidth]{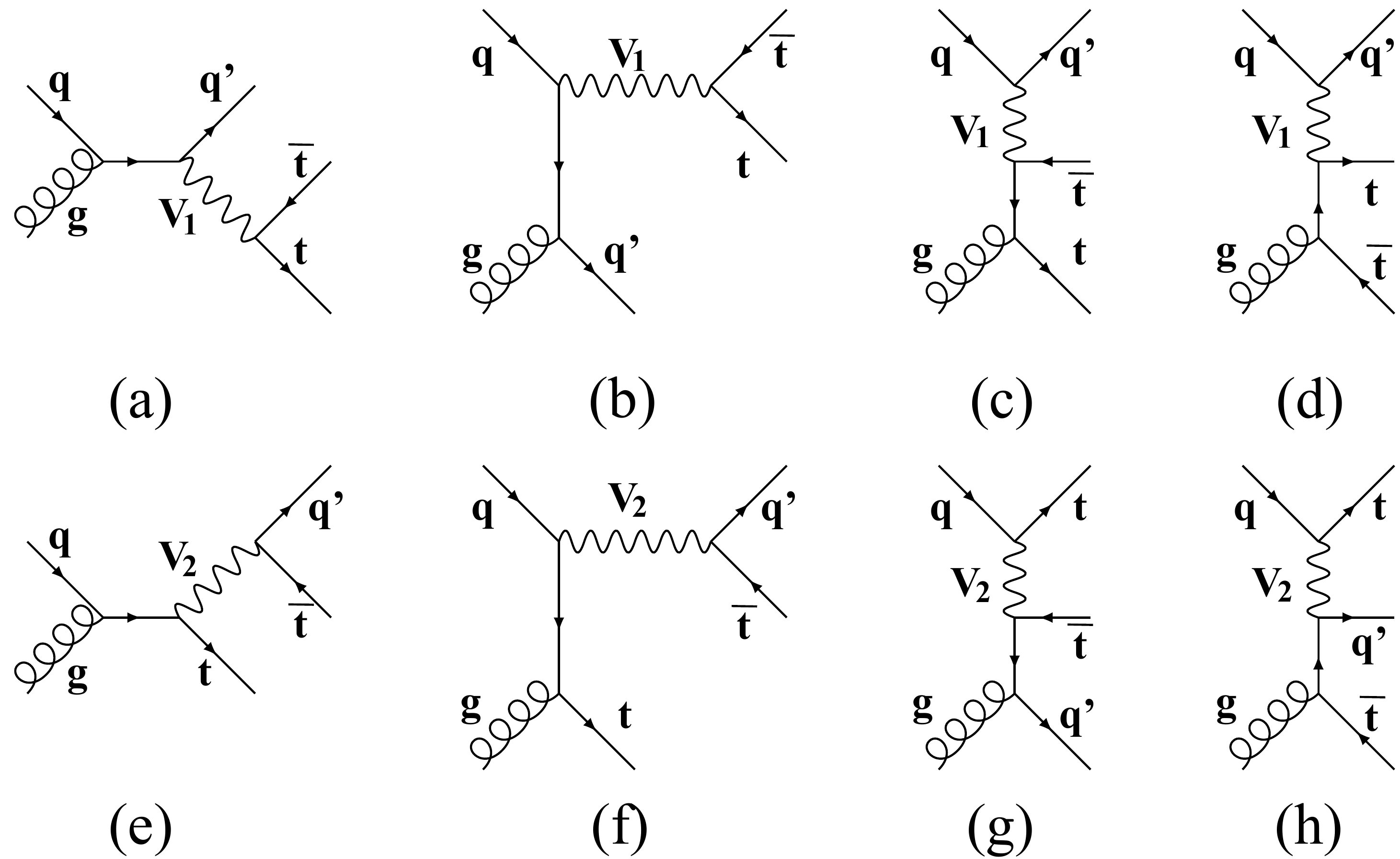}}
\caption{Similar as in Fig.\ \ref{fig:qq-real-pbzpwp-diagrams} but for the subprocesses $g q \to t \bar t q'$.}
\label{fig:qg-real-pbzpwp-diagrams}
\end{figure}

The following $2 \to 3$ tree-level amplitudes, $A_{1/2;1}$, contribute 
to electroweak top-pair production at ${\cal O}(\alpha_s \alpha^2)$:
\begin{inparaenum}[(i)]
\item $q' \bar q \to t \bar t g$ (and the charge conjugated process)
and
\item $g q \to t \bar t q'$ (and the charge conjugated process).
\end{inparaenum}
The corresponding Feynman diagrams are shown in Figs.\ \ref{fig:qq-real-pbzpwp-diagrams} and \ref{fig:qg-real-pbzpwp-diagrams}.

For comparison, the real emission diagrams in Figs.\ 4 and 5 in our previous calculation \cite{Bonciani:2015hgv}
are a (small) subset of the diagrams depicted in Figs.\ \ref{fig:qq-real-pbzpwp-diagrams} and \ref{fig:qg-real-pbzpwp-diagrams}
in a rather compact manner.

\begin{figure}[ht]
\centering
\subfloat{\includegraphics[width=\textwidth]{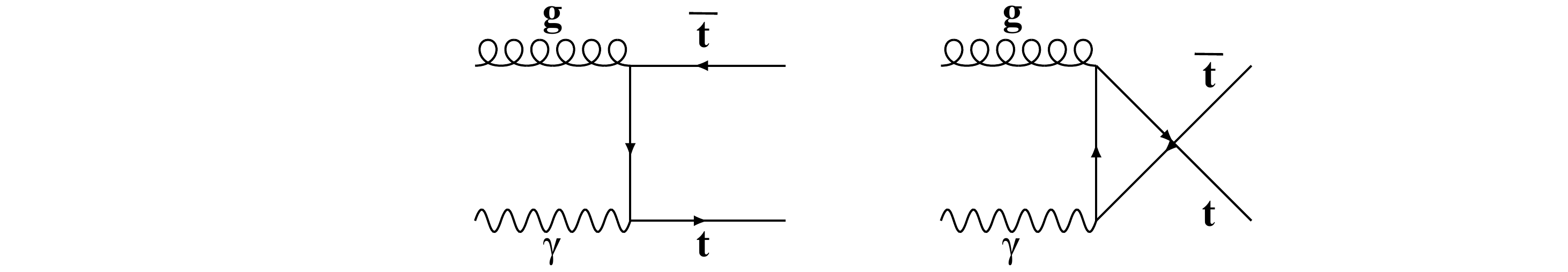}}
\caption{Photon-induced top-pair production of $\mathcal{O}(\alpha_S\alpha)$.}
\label{fig:born-pbzp-diagrams2}
\end{figure}

The $q'\bar q$ subprocesses in Figs.\ \ref{fig:qq-real-pbzpwp-diagrams} contain soft and collinear divergences
which cancel in the sum of real and virtual cross sections as a consequence
of the KLN theorem. As already mentioned, within POWHEG they are treated using the FKS subtraction~\cite{Frixione:1995ms}.
Collinear divergences are present in Figs.\ \ref{fig:qq-real-pbzpwp-diagrams}a), b), e), and f).
On the other hand, collinear gluon emission from a top quark line leads to a finite logarithm of the
top quark mass which we keep in $\hat \sigma$ in fixed order perturbation theory.
The $gq$ and $g\bar q$ channels can only have collinear singularities.
While the diagrams in Figs.\ \ref{fig:qg-real-pbzpwp-diagrams}a), e), f), and h)  are completely finite
the other diagrams in Fig.\ \ref{fig:qg-real-pbzpwp-diagrams} contain configurations
where a light quark propagator [b),g)] or a photon propagator [c),d)] can be close to its mass shell.
As already discussed in \cite{Bonciani:2015hgv},
the fact that the collinear divergences appearing in Figs.\  \ref{fig:qg-real-pbzpwp-diagrams}c) and d) 
involve a photon propagator has two consequences: 
\begin{inparaenum}[(i)]
\item we have to introduce a PDF for the photon inside the proton, 
and 
\item the corresponding underlying Born process shown in 
Fig.\ \ref{fig:born-pbzp-diagrams2}, $g\gamma \to t  \bar t$,
must be included in the calculation.
\end{inparaenum}

\subsection{Validation}
  All the amplitudes with a $Z'$ and $W'$ obtained with {\tt Recola2} have been carefully checked
  to reproduce the SM limit, where we can rely on the correctness of {\tt
  Recola2} having been validated against many different tools such as {\tt
  OpenLoops2} \cite{Buccioni:2019sur} and {\tt MadGraph5\_aMC@NLO}  \cite{Alwall:2014hca}.
Furthermore, the diagonal $\gamma/Z/Z'$ to $t\bar{t}$ amplitudes obtained with {\tt Recola2} have been validated at the amplitude level 
and at the cross-section  level (after implementing them in {\tt POWHEG\,BOX}) at LO and at NLO against our old calculation.
The {\tt POWHEG\,BOX} automatically performs consistency checks on the soft and collinear structure of the amplitudes. 
In addition, we verified analytically the cancellation of soft-real and soft-virtual divergences of our amplitudes.

\section{Models}
\label{sec:models}

With the intention of presenting the different aspects of the new calculation we introduce three different models.

\subsection{Sequential Standard Model}

The Sequential Standard Model (SSM) \cite{Altarelli:1989ff} is a toy model which copies the weak interactions of quarks and leptons by heavier versions $W'$ and $Z'$ of the $W$ and $Z$ boson, respectively. The only  free parameters in this model are the masses of the new heavy gauge bosons. Due to its simplicity and convenience it is a widely used benchmark model in which LHC data are analyzed. The most stringent limits on $W'$ and $Z'$ masses in this model are derived from searches with dilepton final states. Assuming $\Gamma_{Z'}/m_{Z'} =3\%$, a  mass below $5.1$  TeV is excluded by the ATLAS $Z'$ search for high-mass dilepton resonances at the LHC run II with $\sqrt{S}= 13$ TeV and 139 fb$^{-1}$ integrated luminosity (see Fig.\ 3 and Tab.\ 3 of  \cite{Aad:2019fac}). The CMS search for a narrow resonance in high mass dilepton final states using data from LHC run II at $\sqrt{S}= 13$ TeV with 140 fb$^{-1}$ integrated  luminosity leads to a lower mass limit of 
$m_{Z'} \ge 5.15$ TeV assuming a signal width $\Gamma_{Z'}/m_{Z'} = 3\%$ (see Tab.\ 4 of \cite{CMS:2019tbu}). For $W'$ gauge bosons in the SSM, masses below $6.0$ TeV are excluded by 
the ATLAS $W'$ search with charged lepton plus missing transverse momentum final states using  data from LHC run II with  $\sqrt{S}= 13$ TeV and an integrated luminosity of  139 fb$^{-1}$ \cite{Aad:2019wvl}, where $\Gamma_{W'}/m_{W'}$ varies between $2.7\%$ at $m_W' = 150$ GeV and $3.5\%$ above the $t\bar b$ threshold. The CMS $W'$ search using LHC Run II data from 2016 (not the complete Run II data set) at 35.9 fb$^{-1}$ integrated luminosity set the lower limit on the mass of $W'$ to $5.2$ TeV \cite{Sirunyan:2018mpc}.

\subsection{Topcolour model}

The Topcolour (TC) model \cite{HILL1991419, Hill1994hp}  can generate a large top-quark mass through the formation of
a top-quark condensate. This is achieved by introducing a second strong SU(3) gauge group
which couples preferentially to the third generation, while the original SU(3) gauge group
couples only to the first and second generations. To block the formation of a bottom-quark
condensate, a new U(1) gauge group and associated $Z'$ boson are introduced. Different
couplings of the $Z'$ boson to the three fermion generations then define different variants of
the model \cite{Harris1999ya}. A popular choice with the LHC collaborations is the leptophobic TC
model (also called Model IV in the reference cited above) \cite{Harris2011ez}, where the $Z'$ couples only
to the first and third generations of quarks and has no significant couplings to leptons. This particular choice
has three parameters: the ratio of the two U(1) coupling constants, $\cot \theta_H$, which should be large
to enhance the condensation of top quarks, but not bottom quarks, as well as the relative strengths
$f_1$ and $f_2$ of the couplings of right-handed up- and down-type quarks with respect to those of the
left-handed quarks.
This model is excluded by the ATLAS search for $t \bar{t}$ resonances in fully hadronic final states in $p p$ collisions at $\sqrt{S}= 13$ TeV  and an integrated luminosity of  139 fb$^{-1}$  for $Z'$ masses below $3.9$ and $4.7$ TeV and for the decay widths of $1$ and $3\%$, respectively \cite{Aad2020kop}. At $\sqrt{S}= 13$ TeV  and with an integrated luminosity of  35.9 fb$^{-1}$, the CMS search for resonant $t \bar{t}$ production in proton-proton collisions excludes masses up to $3.80$, $5.25$, and $6.65$ TeV for $Z'$ decay  widths of $1$, $10$, and $30\%$, respectively \cite{Sirunyan2018ryr}.

\subsection{Third Family Hypercharge Model}

The Third Family Hypercharge Model (TFHM) \cite{Allanach:2018lvl} is a minimal extension of the SM by an
anomaly-free, spontaneously broken U(1)$_F$ gauge symmetry. Apart from the new gauge boson ($X$) and
a SM singlet, complex scalar field ($\Theta(x)$), needed for the
spontaneous symmetry breaking of the U(1)$_F$ symmetry,  no new particles are introduced. 
The model has flavour-dependent couplings designed to explain various measurements of $B$ meson decays 
($R_K^{(\star)}$ \cite{Aaij:2017vbb,Aaij:2019wad}, 
$BR(B_s \to \mu^+ \mu^-)$ \cite{Aaboud:2018mst,Chatrchyan:2013bka,CMS:2014xfa,Aaij:2017vad}, 
angular distributions in $B\to K^{(\star)} \mu^+ \mu^-$ decays \cite{Aaij:2013qta,Aaij:2015oid,ATLAS:2017dlm,CMS:2017ivg})
which are currently in tension with SM predictions.
In addition, it provides an explanation of the heaviness of the third generation of SM particles and the smallness of the quark mixing. 
An update of the allowed parameter space (post Moriond 2019) can be found in Ref.~\cite{Davighi:2019jwf}.
Recently, the TFHM has been slightly modified to make it more natural in the charged lepton sector \cite{Allanach:2019iiy}. 
In the following we will use the original TFHMeg model from Ref.~\cite{Allanach:2018lvl}. 
The collider phenomenology of the TFHMeg has also been studied in \cite{Allanach:2019mfl}. This model has three free parameters, the extra U(1)
coupling, $g_F$, the angle controlling the mixing of the second and third family quarks, $\theta_{sb}$ and the $Z'$ boson mass. The width of $Z'$ in the TFHMeg is $\Gamma_{Z'} = \frac{5 g_F^2 m_{Z'}}{36 \pi}$. The $Z'$-couplings to the quarks depend on $g_F$ and $\theta_{sb}$ (see Eq.\ (2.15) in \cite{Allanach:2018lvl}):
\begin{equation}
- {\cal L}_{X\psi} = g_F 
\left(\frac{1}{6} \overline{\mathbf{u_L}} \Lambda^{(u_L)} \gamma^\rho \mathbf{u_L}
+ \frac{1}{6} \overline{\mathbf{d_L}} \Lambda^{(d_L)} \gamma^\rho \mathbf{d_L}
+ \frac{2}{3} \overline{t_R} \gamma^\rho t_R 
- \frac{1}{3} \overline{b_R} \gamma^\rho b_R 
\right)Z'_{\rho}\, ,
\end{equation}
where all quark fields are in the mass eigenbasis and $\mathbf{u_L}=(u_L, c_L, t_L)^T$ and $\mathbf{d_L}= (d_L, s_L, b_L)^T$.
The matrix $\Lambda^{(d_L)}$ can be found in Eq.\ (2.16) of Ref.~\cite{Allanach:2018lvl}. It depends on $\theta_{sb}$:
\begin{equation}
\Lambda^{(d_L)} = 
\begin{pmatrix}
0 & 0 & 0
\\
0& \sin^2(\theta_{sb})& \frac{1}{2} \sin( 2 \theta_{sb})
\\
0 & \frac{1}{2} \sin( 2 \theta_{sb}) & \cos^2(\theta_{sb})
\end{pmatrix}\, .
\end{equation}
Moreover,  $\Lambda^{(u_L)}= V \Lambda^{(d_L)}V^\dagger$, where $V$ is the CKM matrix. 
\section{Numerical results}
\label{sec:numerics}

We now use our next-to-leading order calculation to obtain predictions for
top-quark--pair production for the three models introduced in the preceding
section: the Sequential Standard Model, the Top
Colour model, and the Third Family Hypercharge  Model.
Here, our goal is not an exhaustive study of the collider phenomenology for
each of these models scanning over the entire allowed parameter space, but
rather to exemplify our calculation by showing results for a number of
benchmark points.
We will present results for the LHC at $\sqrt{S}=14$ TeV but also for $pp$
collisions at higher centre-of-mass energies.

We will first discuss the general setup of our calculations and event
selection in Secs.~\ref{sec:input} and \ref{sec:cuts} before showing
predictions for fiducial cross sections and NLO $K$-factors in
Sec.~\ref{sec:NLOxsecs}. The impact of the newly included contributions is
discussed in Sec.~\ref{sec:newVsOld}, and finally the impact of the interference
of the BSM signal and the SM background is studied in Sec.~\ref{sec:interf}.

\subsection{Setup and input}
\label{sec:input}

The theoretical description of our calculation and of the models we consider
here can be found in the preceding sections. 
Here we describe the additional input required for the numerical computations
for which the results are presented in the next few sections.
This general setup and the input parameters are used by default if not stated otherwise.

We employ a top quark pole mass $m_t = 172.5$ GeV.
Furthermore, the masses and widths of the weak gauge bosons are given by $m_Z =
91.1876$ GeV, $\Gamma_Z = 2.4952$ GeV, $m_W = 80.385$ GeV, $\Gamma_W = 2.085$
GeV \cite{Patrignani:2016xqp}.
The weak mixing angle is fixed by  $\sin^2 \theta_W = 1-m_W^2/m_Z^2 = 0.222897$
and the fine-structure constant is set to  $\alpha(2 m_t) = 1/126.89$.
We neglect the running of this coupling to higher scales.
We consider quark mixing between all three families and use a unitary CKM
matrix constructed using Wolfenstein parameters as in Ref.~\cite{Zyla:2020zbs}.

For the proton parton distribution functions (PDFs), we use the NLO luxQED set
of NNPDF3.1 \cite{Bertone:2017bme,Manohar:2016nzj,Manohar:2017eqh} as
implemented in the LHAPDF library (ID = 324900)
\cite{Buckley:2014ana,Andersen:2014efa}.
This set provides, in addition to the gluon and quark PDFs, a precise
determination of the photon PDF inside the proton which we need for our cross
section predictions.
The running strong coupling $\alpha_s(\mu_R)$ is evaluated at NLO in the
$\overline{\text{MS}}$ scheme and is provided together with the PDF
set\footnote{Its value is fixed by the condition $\alpha_s(m_Z) = 0.118$.}.

For our numerical predictions in the following sections, we choose equal values
for the factorisation and renormalisation scales, $\mu_F$ and $\mu_R$
respectively, which we identify with  the partonic centre-of-mass energy:
$\mu_F = \mu_R = \sqrt{\hat s}$.
Additionally we vary $\mu_F$ and $\mu_R$ by independently multiplying the
scales by factors of $\xi_R, \xi_F \in \{0.5, 1, 2\}$ discarding combinations with
$\xi_F/\xi_R = 4$ or $1/4$. 
We combine such seven-point variations into an uncertainty band by taking the
envelope of all the predictions.

We present (N)LO+PS predictions for a $pp$ collider with a range of energies
$\sqrt{S} \in \{14, 27, 50, 100\}$ TeV.
We consider the SSM, TC and TFHM models and a range of $Z'$ masses $m_{Z'} \in
[2,8]$ TeV.
In the SSM we set the mass of $W'$ equal to the mass of $Z'$ and its mixing matrix
to that of the SM $W$.
The widths of $Z'$ and $W'$ bosons must then be $\Gamma_{Z'}/m_{Z'} =
3\%$, $\Gamma_{W'}/m_{W'} = 3.3\%$.
The parameters of the TC model are chosen as follows: we set $f_1 = 1$ and $f_2 = 0$
and calculate $\cot \theta_H$ such that $\Gamma_{Z'}/m_{Z'} = 3.1-3.2 \%$.
In the TFHM we set $\theta_{sb} = 0.095$, $g_F/m_Z' = 0.265$ where $m_{Z'}$ is
given in TeV which implies $\Gamma_{Z'}/m_{Z'} = \{0.012, 0.028, 0.050, 0.078,
0.112, 0.152, 0.199\}$ for $m_Z' = \{2, 3, 4, 5, 6, 7, 8\}$ TeV.%
\footnote{This benchmark point was selected from Fig.~1 in \cite{Davighi:2019jwf}.}

\subsection{Event generation and cuts}
\label{sec:cuts}

We generate events in the Les Houches Event format \cite{Alwall:2006yp} using
{\tt POWHEG\,BOX} with stable on-shell top quarks and require the underlying
Born kinematics to satisfy a cut on the $t \bar t$ invariant mass $m_{t \bar
t} \ge 0.75 m_{Z'}$, in order to enhance the signal over background ratio.
We then decay both top quarks leptonically and shower the events using {\tt
PYTHIA 8.244} \cite{Sjostrand:2007gs}.
The branching ratio of the leptonic top decay of $10.5\%$ \cite{Zyla:2020zbs}
squared is applied, unless stated otherwise.
Note that the {\tt PYTHIA} decays wash out any spin correlations.
We use {\tt PowhegHooks} to veto shower emissions harder than the POWHEG
emission and disable QED showers.

We perform further event selection and bin in histograms on-the-fly using {\tt Rivet} \cite{Buckley:2010ar, Bierlich:2019rhm}.
Events are required to have two or more charged leptons, two or more neutrinos,
two or more anti-$k_T$ \cite{Cacciari:2008gp} $R=0.5$ jets each containing at
least one $b$-parton.
All these objects have to fulfil the acceptance cuts $p_T > 25$ GeV and
$|y|<2.5$.
Furthermore, we combine charged leptons and neutrinos into $W$ bosons based on
their MC truth PDG id and require each event to feature at least one such $W^+$
and one such $W^-$ boson.

It is instructive to have a closer look at the size of the various leading order
contributions to the EW top-pair production considered in this study, and the 
effects that the invariant mass and the fiducial cuts have on them.
To that effect in Tab.~\ref{tab:01}, 
\begin{table}
\caption{\label{tab:01}Total cross sections in LO for top-pair production at
	${\cal O}(\alpha_s\alpha)$ and ${\cal O}(\alpha^2)$ in the SM and SSM
	at $\sqrt{S} = 14$ TeV. The $Z'$-boson mass is set to 5 TeV. For all the
	predictions in this table we use NLO $\alpha_S$ and NLO PDFs.}
	\begin{center}
\begin{tabular}{llrr}
	Contribution & no cuts [fb] & $m_{t\bar t}$ cut [fb] & $m_{t\bar t}$ \& fiducial cuts [\%] \\
\hline
	$\gamma g+g\gamma \to t\bar{t},\ {\cal O}(\alpha \alpha_s)$   &   3700   &   0.0327   & 41.6 \\
	$q\bar{q'}\to W \to t\bar{t},\ {\cal O}(\alpha^2)$ + interf.   &   3220   &   0.0573   & 3.7  \\
	$q\bar{q}\to g/W \to t\bar{t},\ {\cal O}(\alpha \alpha_s)$    &   -1680  &   0.000703 & 37.4 \\
	$q\bar{q}\to \gamma/Z\to t\bar{t},\ {\cal O}(\alpha^2)$       &   510    &   0.00614  & 74.9 \\
	$q\bar{q}\to Z' \to t\bar{t},\ {\cal O}(\alpha^2)$            &   0.210  &   0.114    & 77.4 \\
	$q\bar{q'}\to W' \to t\bar{t},\ {\cal O}(\alpha^2)$ + interf.  &   0.0025 &   --       & --   \\
\end{tabular}
\end{center}
\end{table}
we show integrated cross sections in femtobarn for the centre-of-mass energy
$\sqrt{S} = 14$ TeV and the $Z'$ boson mass, $m_{Z'}$, set to 5 TeV with no cuts in
the first column.
The cross sections after the invariant mass cut are shown in the second column and
after both invariant mass and fiducial cuts in the third column.
Note the branching ratio of two leptonic top decays has been stripped from these
predictions.
We do this because the ratio of the first two columns does not depend on the
decay channel and we expect the fiducial cuts, in this study designed for the
dileptonic channel, to have a similar impact in all the other decay channels.
All the contributions in the table are obtained by multiplying an amplitude by
its complex conjugate except for $q\bar{q}\to g/W \to t\bar{t}$, which is
calculated as $M(q\bar{q}\to W \to t\bar{t})M^*(q\bar{q}\to g \to t\bar{t}) +
\text{c.c.}\ $.
The contributions in rows 2 and 6 also contain the interference terms with
the contributions in rows 4 and 5 respectively (indicated by ``+interf.'').

First we observe that the various SM contributions (rows 1-4) are of similar
size, with the resonant production being the smallest.
This must be because the $Z$ boson resonance is below the $2m_t$ threshold.
Furthermore we notice that the ``interference term'' in the third row is
negative, which is not surprising.
The invariant mass cut, see the second column, reduces all the SM contributions by
roughly 5 orders of magnitude, except for the ``interference term'' which is reduced even
more, by about 7 orders of magnitude.\footnote{The invariant mass cut is a
generation cut, so one does not need to worry about the numerical precision in
samples without it.}
The effect of the fiducial cut in the third column is expressed in terms of
percentage relative to the second column.
It has roughly the same impact on all the SM contributions, except for the non-%
resonant $W$ boson production in the second row, in which the bulk of the cross
section is in the forward regions outside the acceptance. 
After both cuts are applied the first two largest contributions are the photon
induced and the resonant $t\bar{t}$ productions, both of which were already
included in our previous calculation \cite{Bonciani:2015hgv}.

By design the invariant mass cut has quite a different impact on the resonant
$Z'$ production and reduces it only gently, by a factor less than two.
As expected the fiducial cut behaves nearly the same for SM and BSM resonant
productions.
After both cuts are applied the $Z'$ contribution is by far the dominant one.
The cuts we designed for this study are thus more than adequate for selecting
SSM $Z' \to t \bar{t}$ production with $m_{Z'} = 5$ TeV at a $\sqrt{S} = 14$
TeV LHC.

The non-resonant $W'$ production is about two orders of magnitude smaller
than the resonant one. 
Moreover we would expect the invariant mass and fiducial cuts to reduce it
considerably similarly to the non-resonant $W$ production.
This contribution in the SSM and at $\sqrt{S} = 14$ TeV is thus
negligible.\footnote{
	At the moment, this contribution cannot be calculated independently of the
	$Z'$ contribution. 
	Because it is much smaller it would require an extremely precise prediction for
	the $Z'$ contribution. 
	Thus we do not report the numbers after cuts.}
Note that this may not be the case anymore at higher collider energies.

\subsection[Fiducial cross sections and NLO $K$-factors]{Fiducial cross sections and NLO \boldmath$K$-factors\label{sec:NLOxsecs}}
\begin{figure}
 \centering
  \includegraphics[width=0.495\textwidth]{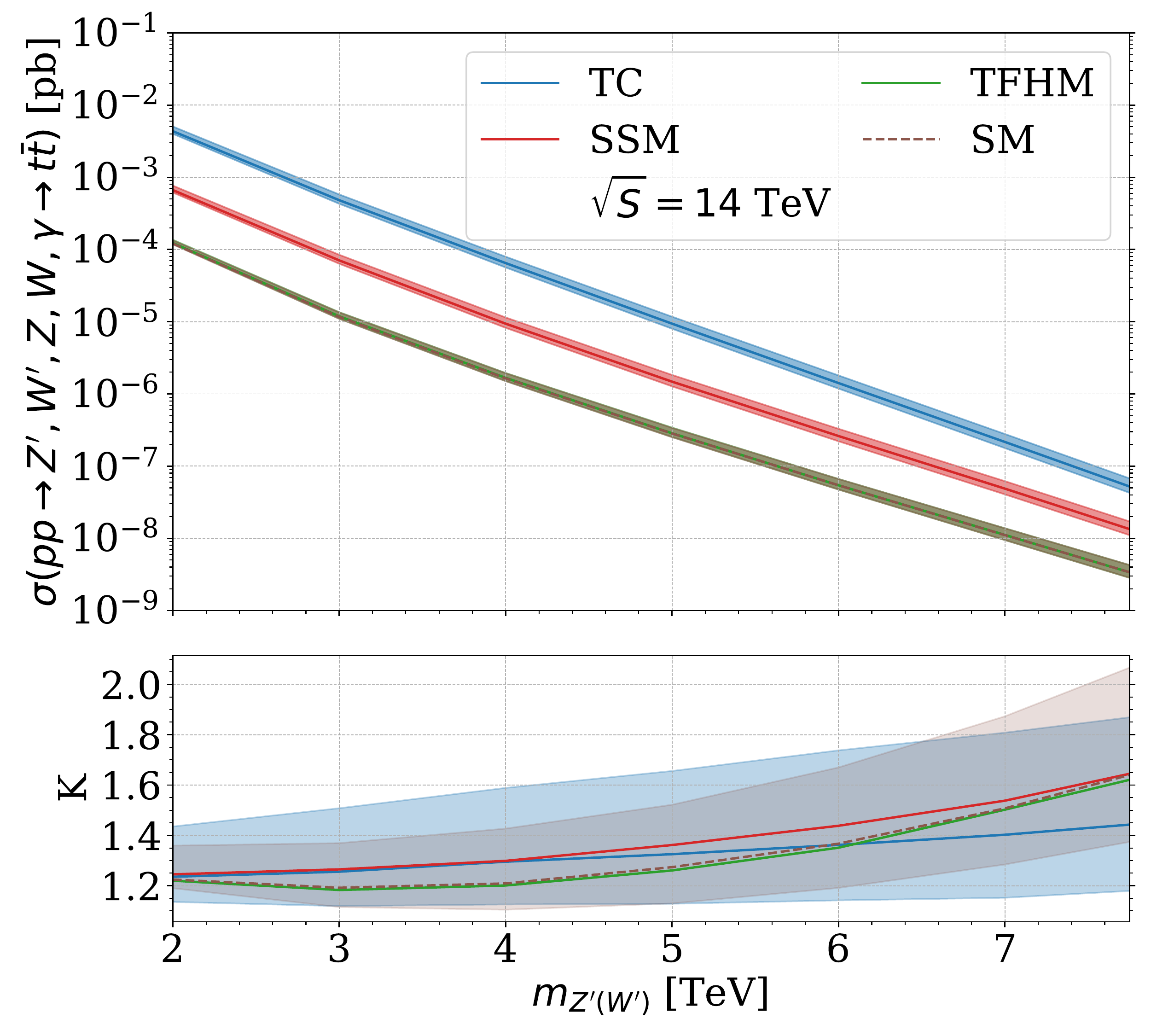}
  \includegraphics[width=0.495\textwidth]{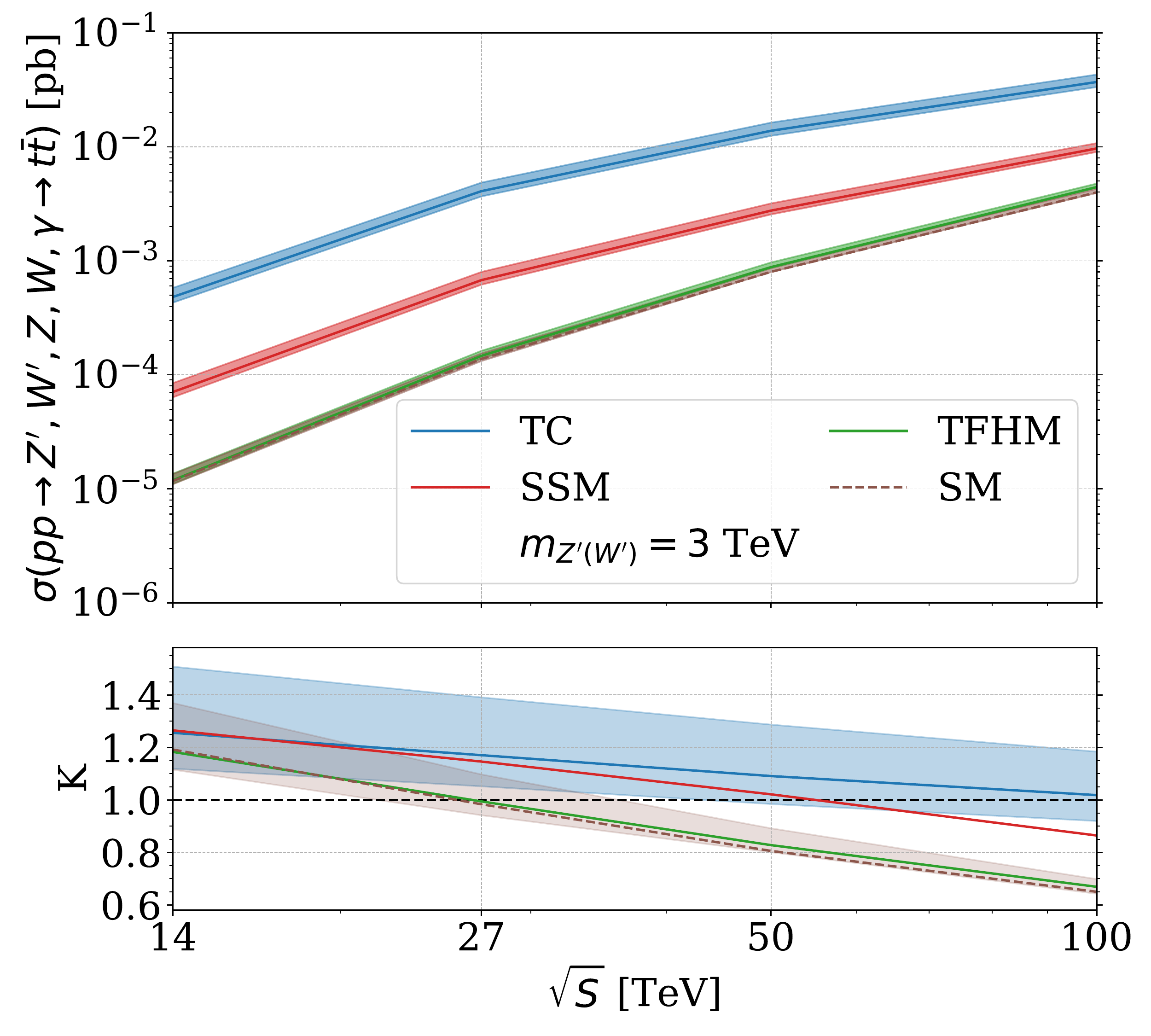} 
\caption{
Fiducial cross sections for EW $t \bar t$ production in the SM, SSM, TC, and
	TFHM with an invariant mass cut $m_{t \bar t} \ge 0.75 m_{Z'}$ and our event
	selection cuts at NLO+PS (upper panels), and as ratio to LO+PS (lower panels).
	In the SSM $m_{W'}=m_{Z'}$. The bands represent perturbative uncertainty due
	to seven-point variation of $\mu_F$ and $\mu_R$.
Left panel: cross sections at $\sqrt{S}=14$ TeV as a function of $m_{Z'}$. 
Right panel: cross sections at $m_{Z'}=3$ TeV as a function of $\sqrt{S}$.}
 \label{fig:01}
\end{figure}

On the upper panels of Fig.~\ref{fig:01} 
we show fiducial NLO+PS cross sections for the SSM, TFHM and the TC model versus
the $Z'$ boson mass, $m_{Z'}$, at a fixed centre-of-mass energy $\sqrt{S}=14$
TeV (left) and versus the centre-of-mass energy for fixed mass $m_{Z'}= 3$ TeV
(right). 
In the SSM the $W'$ boson mass is always set equal to $m_{Z'}$.
For comparison we also include the results for the SM\footnote{Note
that this does not include the QCD contribution.} (grey, dashed line). 
The event generation setup, the invariant mass and the fiducial cuts are as
described above.
In all cases, the cross sections fall off with increasing $m_{Z'}$ and grow
with increasing $\sqrt{S}$.
The former is also true for the SM in which the cross section only depends on
$m_{Z'}$ indirectly through the invariant mass cut.

The invariant mass cut adequately suppresses the SM background relative to
the BSM signal in the SSM and in the TC model.
The prediction for the TFHM model, however, can barely be distinguished from
the SM background throughout the whole mass range at $\sqrt{S}=14$ TeV.
It only becomes appreciably larger than the SM at higher energies, where its ratio
over the SM is roughly 1.13 at $\sqrt{S}=100$ TeV.
Adopting a tighter invariant mass cut would be advised for the TFHM, for
example $m_{Z'} - \Gamma_{Z'} < m_{t\bar{t}} < m_{Z'} + \Gamma_{Z'}$.

The NLO+PS over LO+PS $K$-factors, shown on the lower panels of
Fig.~\ref{fig:01}, are moderate to large and grow with $Z'$ boson mass up to
$\sim 40\%$ in the TC model and up to $\sim 60\%$ in the SSM.
In the absence of BSM effects, this ratio effectively measures the dependence
of higher order corrections on the partonic centre-of-mass energy.
Between 2 and 4 TeV this ratio is fairly flat but then quickly grows, surpassing
$60 \%$ percent at 8 TeV.
As expected, the $K$-factors in the TFHM closely follow those of the SM.

Conversely, the $K$-factors follow the opposite trend versus $\sqrt{S}$ and
eventually almost all drop below one for $m_{Z'} = 3$ TeV.
It would be interesting to see whether the higher order corrections for larger
$Z'$ masses follow a similar pattern.

Higher order corrections are often included in experimental searches in terms
of a constant $K$-factor. 
While this is more or less well justified for a range of $Z'$ masses between 2
and 5 TeV, the corrections more than double when this range is extended to 2-8
TeV.
It may thus be desirable to abandon this crude approximation in high luminosity
or high energy searches where we expect the reach to extend considerably.

\subsection{Impact of non-resonant contributions \label{sec:newVsOld}}

A new feature of our calculation is that we include non-resonant contributions
with $t$-channel $W$, $W'$ and $Z'$ exchange.
We study their impact here.

\begin{figure}
 \centering
 \includegraphics[width=0.495\textwidth]{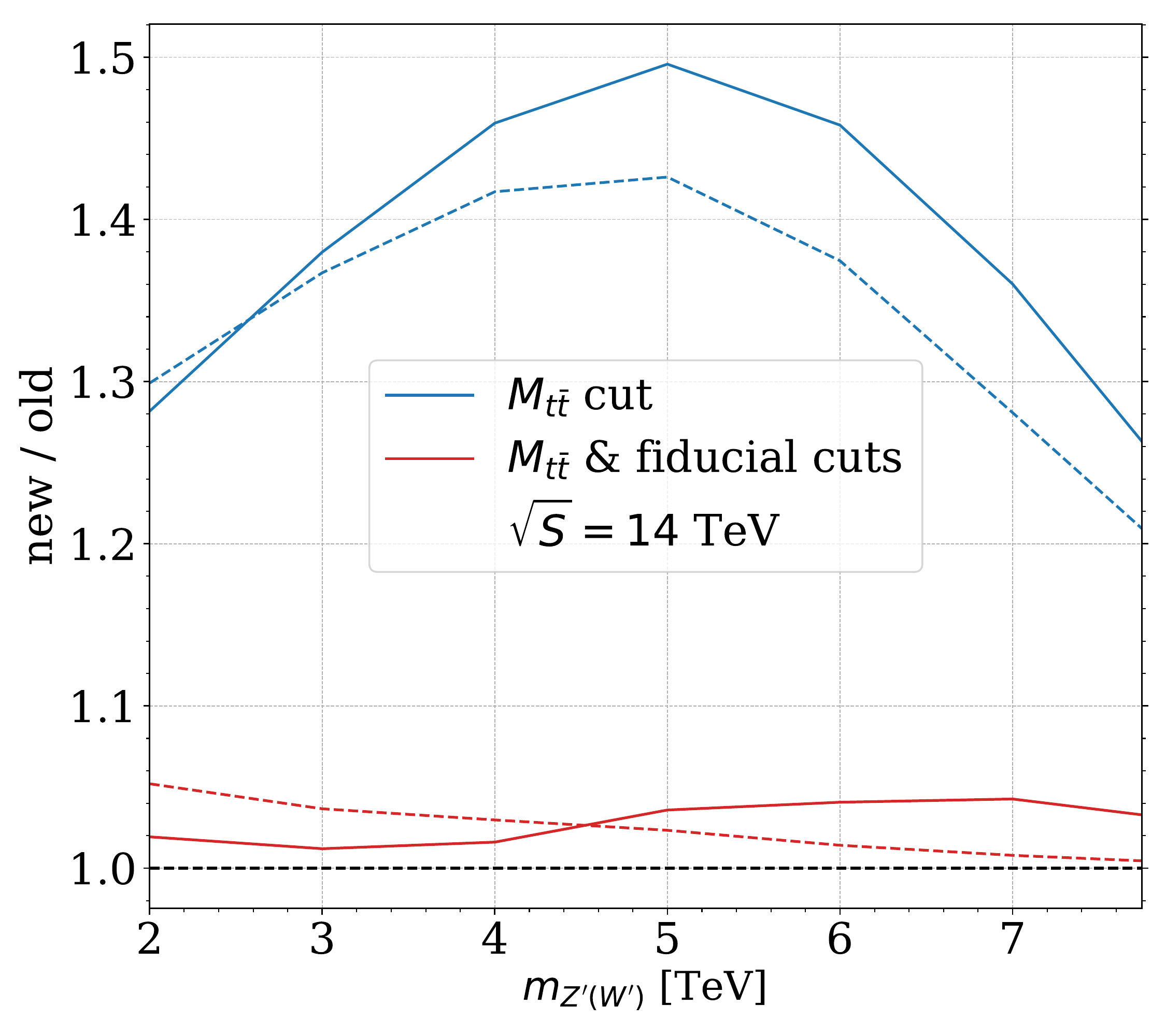}
 \includegraphics[width=0.495\textwidth]{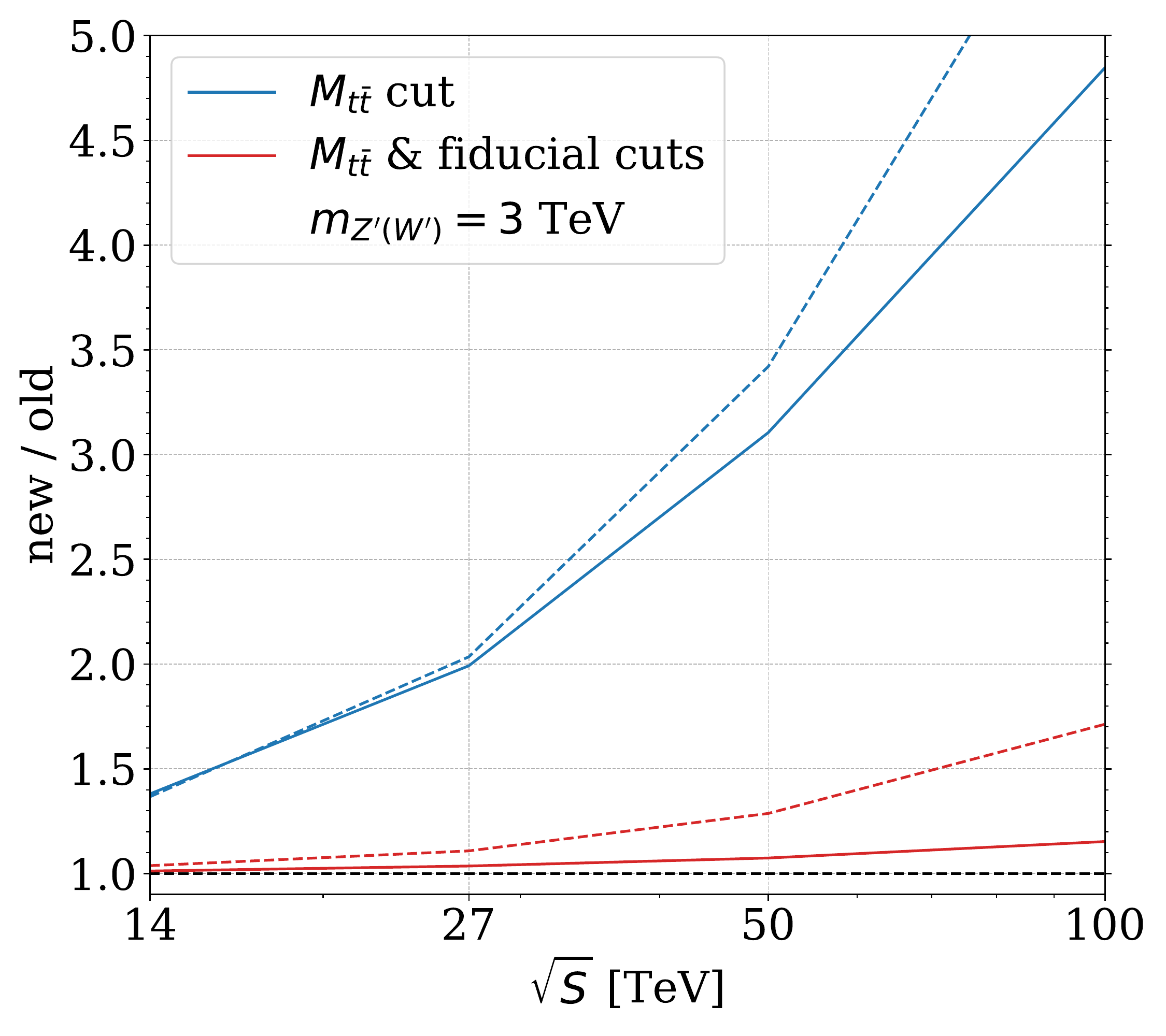}
 \caption{The ratio of cross sections for EW $t\bar t$ production in the SSM
	in the new calculation, including $t$-channel $W$ and $W'$ exchange
	contributions, over the old calculation, without these contributions, at NLO+PS
	(solid lines) and LO+PS (dashed lines). The cross sections have been
	calculated with an invariant mass cut $m_{t \bar t} \ge 0.75 m_{Z'}$ and with
	(red lines) and without (blue lines) fiducial cuts.  Left panel: cross section
	ratio at $\sqrt{S}=14$ TeV as a function of $m_{Z'}=m_{W'}$.  Right panel:
	cross section ratio at $m_{Z'}=m_{W'}=3$ TeV as a function of $\sqrt{S}$.}
 \label{fig:02}
\end{figure}

In Fig.~\ref{fig:02} we show the ratio of predictions for cross sections for EW
$t\bar t$ production in the SSM obtained using our new version of {\tt PBZp}
over the old one of Ref.~\cite{Bonciani:2015hgv}.
The left panel shows this ratio as a function of $m_{Z'}=m_{W'}$ at
$\sqrt{S}=14$ TeV, the right panel as a function of $\sqrt{S}$ at $m_{Z'}=m_{W'}=3$
TeV.
The cross sections have been calculated with an invariant mass cut $m_{t \bar
t} \ge 0.75 m_{Z'}$ and with (red lines) and without (blue lines) fiducial cuts,
at NLO+PS (solid lines) and LO+PS (dashed lines).

The ratio of our predictions at $\sqrt{S}=14$ TeV is roughly between 1.2 and
1.5 when fiducial cuts are not considered.
The reason behind it is the new non-resonant contribution, due to the SM
$q\bar{q}'\to W \to t\bar{t}$ process, not falling off with the invariant mass
as fast as the resonant one. 
This behaviour was already observed in Tab.~1, where this process contributes less than the resonant $Z'$ production
after the invariant mass cut but is still roughly of the same order of magnitude.
In this respect the $q\bar{q}'\to W \to t\bar{t}$ process behaves similarly
to the $\gamma g+g\gamma \to t\bar{t}$ process, which was already included
in our old calculation.
The situation worsens as the centre-of-mass energy is increased, the value
of the ratio reaching almost 5 at $\sqrt{S}=100$ TeV.

Once the fiducial cuts are switched on, the new contributions are reduced
considerably and the ratio of the ``new / old'' predictions drops down to
roughly between 1.0 and 1.05 across the whole mass range at $\sqrt{S}=14$ TeV
and between 1.0 and 1.2 across the whole $\sqrt{S}$ range at fixed mass
$m_{Z'}=m_{W'}=3$.
This is simply because the bulk of the high invariant mass cross section
for the $q\bar{q}'\to W \to t\bar{t}$ process lives in the forward region.
In this respect the $q\bar{q}'\to W \to t\bar{t}$ process is
quite dissimilar to the $\gamma g+g\gamma \to t\bar{t}$ process.

QCD corrections do not change this picture appreciably, but we note that 
the corrections to this ratio can be considerable at high centre-of-mass 
energies, over 50\%.
Our new calculation thus confirms our previous predictions for the SSM at
$\sqrt{S}=14$ TeV, while at the same time it offers a much more sophisticated
description of the interplay of various contributions that enter electroweak
top-pair production.
This interplay may become very important for $Z'$s with weaker couplings and
will certainly become important at higher collider energies.

\subsection{Signal--background interference\label{sec:interf}}
Interferences between the BSM signal and the SM background are routinely
neglected even in the most recent experimental searches.
The argument is that interferences mostly affect the shapes of resonance bumps,
which ``bump-hunting'' is largely insensitive to.
Consequently, experimental analyses work with the SM only and the SM+BSM
hypotheses, where the latter is a ``naive'' sum of the signal and the
background.

While interference effects are expected to integrate out in total cross
sections, they may no longer be negligible once invariant mass and
fiducial cuts are considered.
In this section we explore interference effects by studying ratios of the
fiducial cross sections for EW $t\bar{t}$ production obtained either using
the full process $p p \to \gamma, Z, Z', W' \to t\bar{t}$ or by summing the
SM background process $p p \to \gamma, Z \to t\bar{t}$ and the SSM signal $p p
\to Z', W' \to t\bar{t}$.
Both sets of predictions will include all the contributions to the EW top-pair
production considered in our new calculation with one exception: those obtained
by summing the SM background and the BSM signal will not include any of the
interference terms $\{\gamma,Z,W\} \times \{Z', W'\}$.

\begin{figure}[!h]
 \centering
 \includegraphics[width=0.495\textwidth]{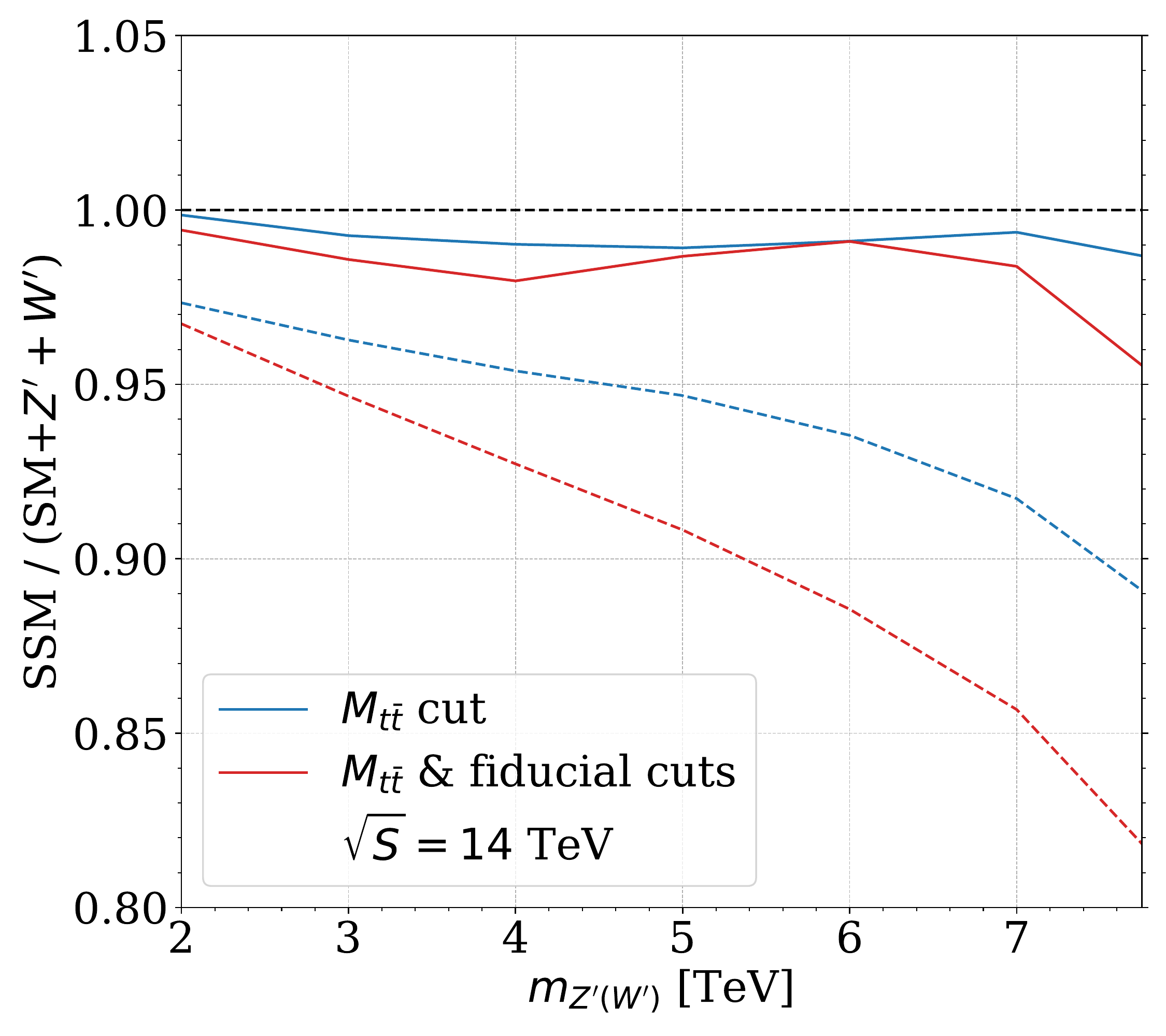}
 \includegraphics[width=0.495\textwidth]{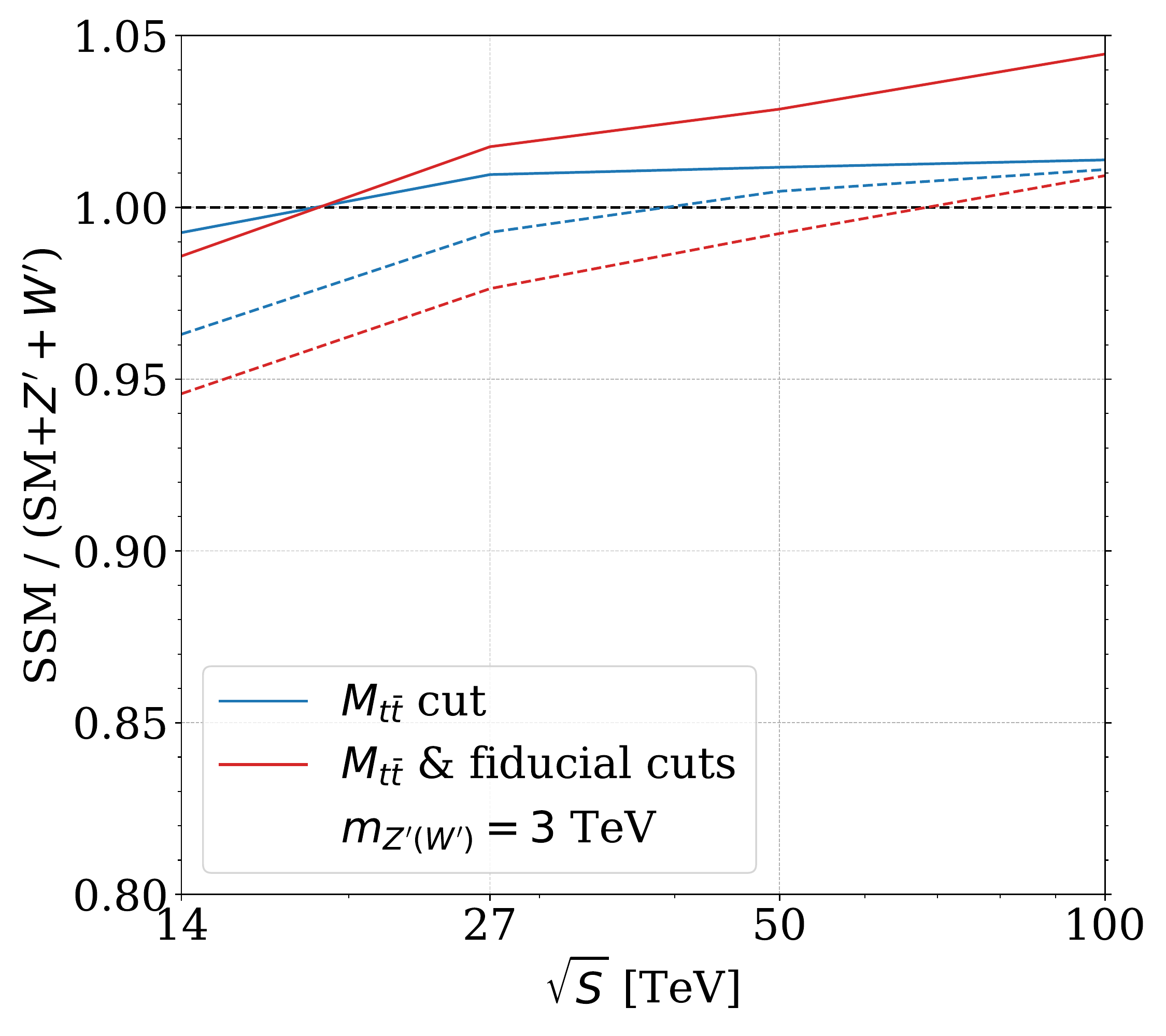}
\caption{The ratio of the cross section for EW $t\bar t$ production in the SSM
	with and without the interference terms between the SM and the $Z', W'$
	contributions at NLO+PS (solid) and LO+PS (dashed).  The cross sections have
	been calculated with an invariant mass cut $m_{t \bar t} \ge 0.75 m_{Z'}$ and
	with (red line) and without (blue line) fiducial cuts.  Left panel: Cross
	section ratio at $\sqrt{S}=14$ TeV as a function of $m_{Z'}=m_{W'}$.  Right
	panel: Cross section ratio at $m_{Z'}=m_{W'}=3$ TeV as a function of
	$\sqrt{S}$.}
 \label{fig:03}
\end{figure}

This ratio of the cross sections with the interference terms over the ones without them
is shown in Fig.~\ref{fig:03}
as a function of $m_{Z'}$ on the left and as a function of $\sqrt{S}$ on the
right. 
As in the previous section, the predictions with the invariant mass cut are
shown in blue, the predictions with both the invariant mass and the fiducial
cuts in red; the NLO+PS predictions are plotted with solid lines, while LO+PS
ones with dashed lines.
We find that the interference reduces the cross section at $\sqrt{S} = 14$ TeV
and has a relatively steep profile versus $m_{Z'}$ at LO: a few percent for
the light $Z'$s to well over 20\% for the heavy ones.
The size of the interference effects seem rather flat as a function of
$\sqrt{S}$ for fixed $m_{Z'} = m_{W'} = 3$ TeV in comparison.
We also observe that the interference effects tend to be pronounced by
the fiducial cuts, while the higher order corrections rather stabilise them.

Note that these conclusions may not generalise, as we expect the interference
effects to strongly depend on the BSM scenario and on the position of the
invariant mass cut.
In view of their potentially large size, however, we advocate they be
considered in experimental searches.

\section{Summary and conclusions}
\label{sec:conclusions}

We extended and improved upon our previous calculation 
of electroweak top-quark pair hadroproduction in extensions
of the Standard Model with extra heavy neutral and charged spin-1 
resonances.
In particular, we now allow for flavour-non-diagonal $Z'$ couplings in order to accommodate
a wider class of heavy resonance models including models which have been brought forward
to explain the anomalies in $B$ decays.
We now also take into account non-resonant production 
in the SM and beyond, including the contributions with $t$-channel $W$, $W'$  and $Z'$ bosons. 
Compared to our previous work, the entire chain of tools used for the calculation changed;
all amplitudes are now generated using the {\tt Recola2} package. Our calculation is one of the first
to use {\tt Recola2} for a BSM calculation.
As in our previous work, we included NLO QCD corrections and consistently matched them to parton showers with the 
POWHEG method fully taking into account the interference effects between SM and new physics amplitudes. 
This study paves the way for a similar upcoming calculation for the $t\bar b$ final state.

As a first application, we presented numerical results for $t \bar t$ cross sections at hadron colliders with a centre-of-mass energy up to 100 TeV
for three models, the Sequential Standard Model, the Topcolour model, as well as the Third Family Hypercharge Model leaving
a more detailed analysis including comparisons with LHC data for a future study. 
We discussed the effect of cuts on the signal over background ratio and present $K$-factors which turned out to increase considerably as a function of the heavy resonance mass. The impact of the new contributions
was shown to be modest at 14 TeV if suitable cuts are applied. However, they are expected to become sizable at a Future Circular Collider operated
at 100 TeV. At such energies it would be interesting to compare our predictions with results obtained in a 6-Flavour Number Scheme including 
parton densities for the top quark and the weak bosons of the Standard Model.

\subsection*{Acknowledgments}
The work of T.J. was in part supported by the DFG under grant 396021762 -
TRR257 and in part by the SNSF under contracts BSCGI0-157722 and
CRSII2-160814. The work of M.M.A. and I.S. was supported in part by the IN2P3 master project “Th\'eorie – BSMGA”. M.M.A. is thankful for the hospitality of the Physics
Institute of University of Z\"urich where part of this work was
performed. J.-N.L. was supported by the SNSF under contract
BSCGI0-157722. This work has also been supported by the BMBF under
contract 05H18PMCC1.

\bibliographystyle{JHEP}
\bibliography{pbvp}

\end{document}